# A new kind of vortex pinning in superconductor / ferromagnet nanocomposites

A. Palau,[1] H. Parvaneh,[1] N.A. Stelmashenko,[1] H. Wang[2] J. L. Macmanus-Driscoll[1] and M.G. Blamire[1]

[1] Department of Materials Science, University of Cambridge, Pembroke Street, Cambridge CB2 3QZ, U.K.

[2] Department of Electrical and Computer Engineering, Texas A & M University, College Station, TX 77843-3128

This paper reports the observation of hysteresis in the vortex pinning in a superconductor / ferromagnetic epitaxial nanocomposite consisting of fcc Gd particles incorporated in a Nb matrix. We show that this hysteretic pinning is associated with magnetic reversal losses in the Gd particles and is fundamentally different in origin to pinning interactions previously observed for ferromagnetic particles or other microstructural features.

There has been considerable recent interest in the properties of coupled superconductor (S) / ferromagnet (SF) systems [1]. In addition to interactions at the electronic level [2] the F magnetic moment can directly influence the superconducting properties [3,4]. For example, in reverse domain superconductivity (RDS) [5], fields originating from F particles locally oppose an applied field and shift the magnetic phase diagram so that the maximum critical temperature ($T_c$) occurs at non-zero field. RDS has been observed in both bilayer films [5] and in S layers superimposed on arrays of ferromagnetic dots [4].

Zero-voltage currently flow in a superconductor is limited by the ability of the material to pin quantized flux vortices on microstructural features and so prevent their flow under the Lorentz Force (LF) given by **J**×**B**. Numerous studies have investigated the potential for strong pinning through the incorporation of F particles within a S matrix [6-8]: although some pinning enhancements have been observed, the technique is not applied commercially and the nature of the pinning interaction is not well-understood, partly because of the difficulty of achieving a sufficiently fine dispersion of the appropriate particles. Our aim in this research was to create a self-assembled SF nanocomposite system by exploiting the mutual insolubility [7] of Nb (S) and Gd (F).

In this Letter we report measurements on Nb/Gd films which show enhanced vortex pinning only for decreasing fields. We show that this pinning cannot be explained in terms of the conventional mechanisms of "core pinning" based on position-variation in the condensation energy, or "magnetic pinning" associated with disturbance of the vortex screening currents by extended defects. Instead we introduce a new pinning mechanism based on hysteresis losses in the F component of the composite.

In order to obtain a fine dispersion of ferromagnetic particles, the elements were co-deposited at high temperature so that surface spinodal decomposition largely controlled the mean Gd particle size; a similar approach has been used to create W/Gd nanocomposite films [9]. Films were grown in an ultra-high vacuum system equipped with two d.c. sputter sources onto r-plane (1$\bar{1}$02) sapphire; routine conditions for epitaxial growth of Nb (750°C and 0.1 nm s$^{-1}$) were used. Early trials demonstrated the tendency for Gd to react with the substrate at higher temperatures and therefore, although a lower temperature was finally used, we followed Surgers *et al.* [10] and grew a Nb buffer layer as a precaution against reaction. The typical buffer thickness was less than 50 nm, with the total film thickness of the order of 200 nm. The composition of each sample was measured by energy dispersive X-ray (EDX) analysis using the Nb L and Gd M lines; the composition of the samples reported in this paper was between 20 and 40 atomic percent Gd.

X-ray diffraction showed that in all cases the Nb was predominantly (001) oriented with a typical rocking curve width of the (002) peak of ~0.4°, indicative of good crystallographic alignment. Reliable identification of the Gd peaks was much more difficult: a number of weak peaks, corresponding to both cubic (fcc) and hexagonal phases, were observed. The weakness of the peaks is consistent with the Gd crystallite size deduced from high resolution transmission electron microscopy (TEM) as shown in Fig. 1; the scale of the lattice distortions associated with the Gd inclusions (marked with arrows) within the Nb lattice suggests a typical Gd crystallite size of <10 nm. Larger area images showed good alignment of the Nb and sapphire lattice images across the substrate interface. The selected area diffraction pattern (inset to Fig. 1) is most readily interpreted as originating from fcc Gd within Nb: the lattice orientations between each materials are determined to be: zone orientation: Nb<010> // Gd<110>

// $Al_2O_3$<10$\bar{1}$2>, plane orientation (200)Nb // (220)Gd // (01$\bar{1}$2) sapphire; the diffraction pattern of Gd is not consistent with hcp bulk Gd. Thus the Gd structure and orientation in the epitaxial nanocomposite is different to that observed previously in epitaxial multilayers [11].

Figure 2 shows a typical magnetization vs field loop measured using a vibrating sample magnetometer just above $T_c$. The hysteresis implies that at least a proportion of the Gd is ferromagnetic although the effective moment per Gd atom is only about 10% of the bulk value. In the superconducting state this moment is superimposed on the hysteresis arising from flux pinning within the films. The $T_c$ (~8.5 K), measured using a four point transport probe, varied weakly over the composition range of the samples reported in this paper, but decreased very rapidly to below 4.2 K for Gd compositions exceeding 40 at%. The residual resistance ratio measured at 10 K of 10 translates via the usual relation, $l = m^* v_F / n e^2 \rho_0$ where $m^*$ is the electron effective mass, $v_F$ is the Fermi velocity, $n$ is the electron density and $e$ the electronic charge, to a low temperature mean free path ($l$) of 30 nm. The upper critical field ($B_{c2}$) was close to the bulk value (0.35 T) for Nb

Critical current vs magnetic field ($J_c$ vs $H$) measurements were performed in a variable temperature cryostat; in all cases the curves show a marked hysteresis in which the curve maxima are shifted from zero depending on the direction of the field sweep, with the maximum $J_c$ measured during decreasing field measurements following field-cooling (FC) through $T_c$. Fig. 3 shows the temperature evolution of this hysteresis. This hysteresis was absent in measurements of control samples of pure Nb deposited under the same conditions.

At first sight, despite the epitaxial nature of the films, the $J_c$ vs $H$ hysteresis appears similar to that measured in granular $YBa_2Cu_3O_{7-d}$ [12,13] which can be explained on the basis of reverse fields due to trapped current loops within the material – the direct superconducting analogy of RDS. The key difference observed in our measurements is that, unlike the hysteresis in granular materials (for example [12,14]), $J_{cmax}$(FC) > $J_{cmax}$ (ZFC). The reason for this is straightforward; as in the case of RDS, the effect of the local reverse fields in granular systems is partially to cancel the applied field and so restore a *fraction* of the material to the zero-field maximum $J_c$; thus the total $J_c$ cannot exceed $J_c$(ZFC). In addition, the total remanent moment of the Gd particles is such that the reverse field which, by analogy with the thin film magnetic dot arrays [4], might be associated with any particle clusters is negligibly small (<$10^3$ A m$^{-1}$).

It is worth noting at the outset that, in our experiment, direct coupling between vortices and individual Gd nanoparticles is precluded by the particle size and density which is significantly smaller than both the penetration depth $\lambda$ and the coherence length $\xi$ in Nb. Indeed the particle density and size are such for Nb the volumes $\lambda^3$ and $\xi^3$ will contain hundreds of 5-10 nm diameter particles (see inset (a) of Fig. 2) and in pinning terms the material might behave more like a continuum magnetic superconductor. In contrast, in earlier experiments [7,8] the particle density was such that interaction between vortices and individual particles was inevitable. In addition, in all previous experiments it has been assumed (explicitly or otherwise) that the ferromagnetic moment is not perturbed by changes in the superconducting state. It is these differences in our experimental system which give rise to the new pinning mechanism.

At low fields in a type II superconductor, confinement of the magnetic flux within a vortex screens the local magnetic field from the bulk of the superconductor. Thus in the Nb/Gd nanocomposite system the field experienced by a Gd particle will be higher if it lies within the vortex than outside. This means that vortex displacement will change the effective field on the Gd particles at both the new site and the old (see insets to Fig. 4). Any hysteresis in the magnetization of the nanoparticles will mean that this displacement requires energy input. We will show that this new form of flux pinning associated with magnetic hysteresis can account for all the features of our experimental results.

We consider a vortex to be represented by a cylindrical flux bundle of radius $\lambda$, and so the magnetic (Zeeman) energy per unit length of vortex is $\pi\lambda^2 B_V M$ where $B_V$ is the flux density within the vortex and $M$ is the mean magnetization of the composite within the vortex. We start by assuming that a field sufficient both to saturate the Gd and exceed $H_{c2}$ is applied and then progressively reduced. At any field, a vortex moving from one position to another reduces the magnetization at the original core location from $M_1$ to $M_2$ (see the schematic magnetization plot inset (b) of Fig. 2) and increases the magnetization at the new location from $M_2$ to $M_3$ (since the field at the new core location will already have been reduced to zero). Therefore the change in energy per unit length for a vortex moving from site A to site B is given by

$$\Delta\varepsilon = \pi\lambda^2 B_V (M_3 - M_1) \qquad (1)$$

This can be related to the pinning force per unit length by:

$$\Delta J_c \Phi_0 = f_p \simeq \Delta\varepsilon / \lambda = \pi\lambda B_V (M_3 - M_1) \qquad (2)$$

where $\Delta I_c$ is the critical current enhancement, which is therefore given by:

$$\Delta J_c \simeq \pi\lambda B_V (M_3 - M_1) / \Phi_0 \qquad (3)$$

The analysis presented applies for fields reduced to zero from a high-field starting position. Continuing to reduce the field through zero will introduce vortices of opposite polarity which will induce a magnetization $M_4$ at their core. Displacement of these vortices will induce the same magnetization $M_4$ at the new site and therefore the magnetic energy is independent of position and the excess pinning due to magnetic hysteresis will have vanished. The same is true in the case of increasing field from zero, in which case the magnetization at the vortex core is always $M_3$. Thus magnetic hysteresis pinning is expected to result in an enhanced $J_c$ on reducing the field from $H_{c2}$, but will vanish once zero field is reached. Its signature is a true enhancement of pinning, rather than a shifted field origin. Residual magnetic pinning will remain which is associated with the minor hysteresis loop between $M_2$ and $M_3$ (see inset (b) of Fig. 2), but this can be expected to be a significantly smaller effect. Since it is entirely magnetic in origin, the magnitude of $f_p$ for magnetic hysteresis pinning will depend on the inhomogeneity of the field in the mixed state. Therefore it is only expected to be significant for flux densities $B_{hyst}$ given by $\Phi_0/B_{hyst} \geq \sqrt{3}a_0^2/2$ where $a_0 \sim 2\lambda$. Taking $\lambda(T) = \lambda_0\left\{1-(T/T_c)^4\right\}^{-1/2}$ implies that a plot of $B_{hyst}$ vs $1-(T/T_c)^4$ should have a gradient of $\Phi_0/2\sqrt{3}\lambda_0^2$. This is shown in Fig. 4 from which $\lambda_0 = 107\pm10$ nm, consistent with the normal value for thin film Nb of about 120 nm [15], but which can in the clean limit this can decrease to ~50 nm [16].

Using this value, the core flux density for a vortex in our films at 5 K is $B_V = \Phi_0/\pi\lambda^2 = 56$ mT. We do not know the precise shape of the minor loop, but we estimate from Fig. 2 that the change in moment corresponding to $M_1$-$M_3$ is $\sim 10^{-8}$ Am$^2$. Given the sample dimensions this translates to 2000 A m$^{-1}$. Substituting the appropriate values into equation (3) gives $\Delta J_c \sim \times 10^{10}$ A m$^{-2}$: this is similar to the enhanced FC pinning shown in Fig. 3, but in practice it may be enhanced by any inhomogeneity in the nanoparticle density.

Several aspects of this analysis require further comment.

Firstly, we have used the measured average moment of the composite to calculate the pinning force. It is clear the Gd moment in our particles is substantially suppressed compared to the bulk. While this is possibly due to surface interactions (the magnetic "dead-layer" of > 0.5 nm routinely observed in S/F multilayer structures [17]) or finite-size effects [9,11], there is considerable uncertainty about the magnetic properties of the cubic Gd phase [18]. In any case, improved engineering of the particle size and distribution might be expected to enhance $M$ and hence $J_c$ by a large factor.

Secondly, it is reasonable to suppose that a 5-10nm Gd particle might be single-domain and, since there seems to be significant crystallographic texture, it might also be concluded that the particle easy axes should be aligned and hence complicates the analysis. From the Fig. 2 it is clear that the composite film does not coherently reverse and so we have chosen to model the system as if it were a continuum magnetic superconductor with a magnetic hysteresis loop. The likelihood is that there is a range of sizes, shapes and orientations of the Gd particles within our samples which will have the effect of creating a distribution of easy axis directions and particle coercivities which will behave collectively as measured. In addition, it seems likely that at the high particle densities considered in this work there could be a considerable dipolar or RKKY coupling [19] between them.

Thirdly, the pinning arises from magnetic hysteresis losses associated with vortex displacement and therefore requires that the fields associated with vortices are sufficient to partially reverse the ferromagnetic particles. This is the opposite of the Fe/Hg-In case [8] in which the Fe moment was not altered by field-cycling the superconductor. In the latter case a uniform *high density* of particles would yield little enhancement in pinning as the magnetic energy would be independent of vortex position. In fact, although our experimental system is an S/F nanocomposite, the proposed pinning mechanism applies most naturally to intrinsic ferromagnetic superconductors. Interestingly Gammel *et al.* [20] demonstrate a substantially increase in pinning when $ErNi_2B_2C$ is cooled through a ferromagnetic transition temperature below $T_c$ which may relate to this mechanism.

In conclusion, our experimental results can be explained by a form of flux pinning which depends on the irreversible magnetization within a homogeneous (on the scale of $\lambda$) ferromagnetic component within a type II superconductor. This mechanism is different to standard pinning interactions and may have more general relevance.

The authors would like to acknowledge helpful discussions with A.M. Campbell, and Z.H. Barber. Preliminary work on the Nb/Gd system was performed by E.J. Rees. The work was partially support by U.K. EPSRC.

[1]  A. I. Buzdin, Rev. Mod. Phys. 77, 935 (2005); I. F. Lyuksyutov and V. L. Pokrovsky, Adv. Phys. 54, 67 (2005).

[2]  V. V. Ryazanov, V. A. Oboznov, A. Y. Rusanov, A. V. Veretennikov, A. A. Golubov, and J. Aarts, Phys. Rev. Lett. 86, 2427 (2001); T. Kontos, M. Aprili, J. Lesueur, F. Genet, B. Stephanidis, and R. Boursier, Phys. Rev. Lett. 89, 137007 (2002); A. Y. Rusanov, S. Habraken, and J. Aarts, Phys. Rev. B 73, 060505 (2006); A. J. Drew, et al., Phys. Rev. Lett. 95, 197201 (2005); A. Y. Rusanov, M. Hesselberth, J. Aarts, and A. I. Buzdin, Phys. Rev. Lett. 93, 057002 (2004); R. J. Kinsey, G. Burnell, and M. G. Blamire, IEEE Trans. Appl. Supercon. 11, 904 (2001).

[3]  I. F. Lyuksyutov and V. Pokrovsky, Phys. Rev. Lett. 81, 2344 (1998).

[4]  M. Lange, M. J. Van Bael, Y. Bruynseraede, and V. V. Moshchalkov, Phys. Rev. Lett. 90, 197006 (2003).

[5]  J. Fritzsche, V. V. Moshchalkov, H. Eitel, D. Koelle, R. Kleiner, and R. Szymczak, Phys. Rev. Lett. 96, 247003 (2006).

[6]  N. D. Rizzo, J. Q. Wang, D. E. Prober, L. R. Motowidlo, and B. A. Zeitlin, Appl. Phys. Lett. 69, 2285 (1996); A. Snezhko, T. Prozorov, and R. Prozorov, Phys. Rev. B 71, 024527 (2005); M. J. Van Bael, K. Temst, V. V. Moshchalkov, and Y. Bruynseraede, Phys. Rev. B 59, 14674 (1999).

[7]  C. C. Koch and G. R. Love, J. Appl. Phys. 40, 3582 (1969).

[8]  T. H. Alden and J. D. Livingston, J. Appl. Phys. 37, 3551 (1966).

[9]  C. E. Krill, F. Merzoug, W. Krauss, and R. Birringer, Nanostructured Materials 9, 455 (1997).

[10]  C. Surgers and H. Vonlohneysen, Thin Sol. Films 219, 69 (1992).

[11]  U. Paschen, C. Surgers, and H. Vonlohneysen, Z. Phys. B 90, 289 (1993).

[12]  J. E. Evetts and B. A. Glowacki, Cryogenics 28, 641 (1988).

[13]  A. Palau, et al., Appl. Phys. Lett. 84, 230 (2004).

[14]  M. N. Kunchur and T. R. Askew, J. Appl. Phys. 84, 6763 (1998); P. Mune, F. C. Fonseca, R. Muccillo, and R. F. Jardim, Physica C 390, 363 (2003).

[15]  H. Zhang, J. W. Lynn, C. F. Majkrzak, S. K. Satija, J. H. Kang, and X. D. Wu, Phys. Rev. B 52, 10395 (1995).

[16]  C. E. Cunningham, G. S. Park, B. Cabrera, and M. E. Huber, Appl. Phys. Lett. 62, 2122 (1993).

[17]  J. W. A. Robinson, S. Piano, G. Burnell, and M. G. Blamire, Phys. Rev. Lett. in press (2006); J. S. Jiang and C. L. Chien, J. Appl. Phys. 79, 5615 (1996).

[18]  O. Singh and A. E. Curzon, Sol. St. Commun. 44, 1121 (1982); P. E. Chizhov, A. N. Kostigov, and V. I. Petinov, Sol. St. Commun. 42, 323 (1982).

[19]  S. P. Kruchinin, Y. I. Dzezherya, and J. F. Annett, Supercond. Sci. Tech. 19, 381 (2006).

[20]  P. L. Gammel, B. Barber, D. Lopez, A. P. Ramirez, D. J. Bishop, S. L. Bud'ko, and P. C. Canfield, Phys. Rev. Lett. 84, 2497 (2000).

Figure Captions

FIG. 1. Cross-sectional high resolution transmission electron microscopy image of a 30%Gd sample. Gd inclusions are marked as arrows. Inset; selected area diffraction pattern from the same sample.

FIG. 2. (Color online) Magnetic moment vs field at 10 K for Nb/ 30% Gd sample. Insets (a), schematic diagram of a vortex enclosing a number of small magnetic particles; (b) schematic diagram of the magnetization changes experienced during vortex displacement, particles within the vortex (colored points) and outside the vortex (black point).

FIG. 3. (Color online) critical current density vs applied field for a 20%Gd sample for different temperatures: top to bottom 5 K, 6 K, 7 K, 8 K, 8.5 K; closed symbols – decreasing field, open symbols – increasing field, triangles – zero field cooled (ZFC).

FIG. 4. (Color online) Plot of the hysteresis onset field vs $1-(T/T_c)^4$ for the data from Fig. 3 with $T_c$=8.6 K; the dashed line shows a linear fit corresponding to a gradient $\Phi_0/2\sqrt{3}\lambda_0^2$ with $\lambda_0$ = 107 nm. Inset (a) schematic diagram of the interaction between the moments of an array of ferromagnetic particles and the field within a flux vortex; (b) the changes induced in the particle moments by movement of the vortex from site A to site B.

FIG 1.

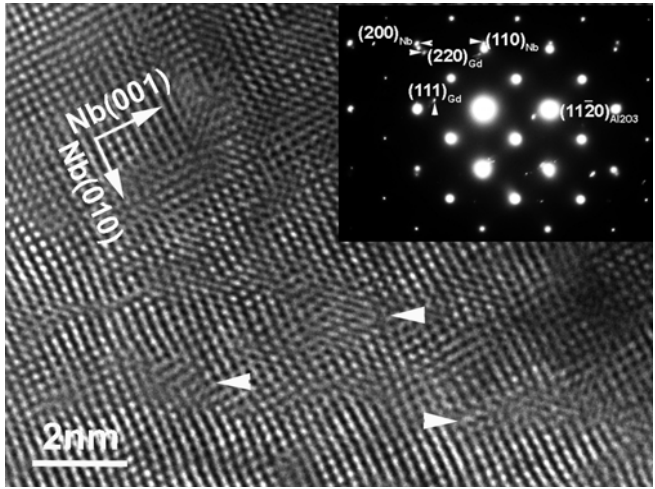

FIG 2.

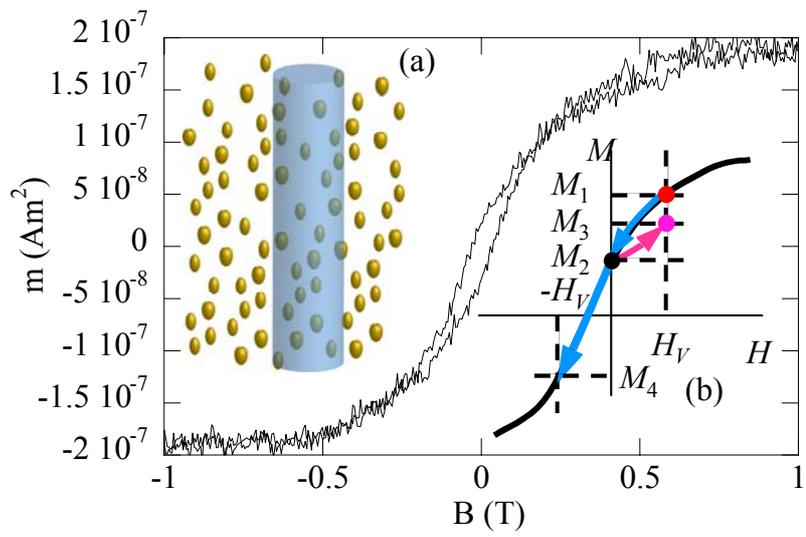

FIG 3.

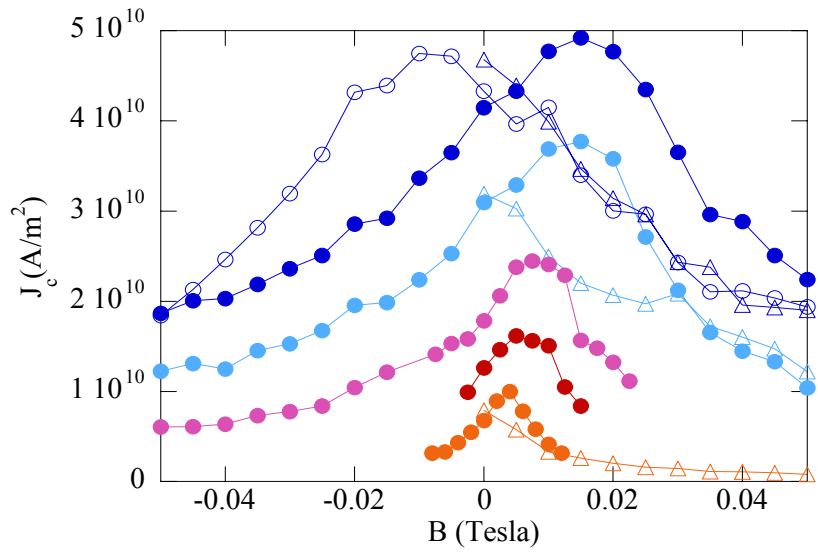

FIG 4.

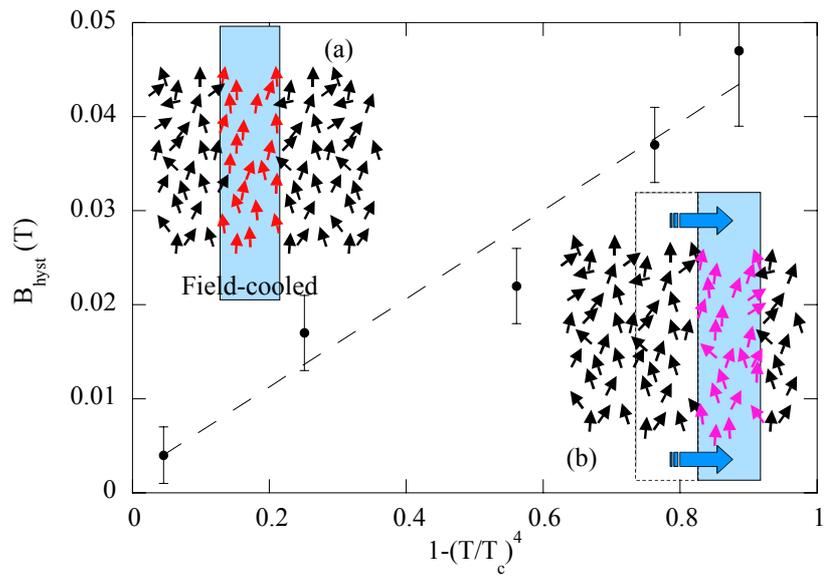